\documentclass[amsmath,amssymb,twocolumn,showpacs]{revtex4}

\usepackage{graphicx}
\usepackage[usenames]{color}

\begin{document}

\title{Raman scattering evidence for a cascade-like evolution of the charge-density-wave collective amplitude mode}
\author{M. Lavagnini$^{1}$, H.-M. Eiter$^{2}$, L. Tassini$^{2}$, B. Muschler$^2$, R. Hackl$^2$, R. Monnier$^1$, J.-H. Chu$^3$, I.R. Fisher$^3$ and
L. Degiorgi$^{1}$} 
\affiliation{$^1$Laboratorium f\"ur Festk\"orperphysik, ETH -
Z\"urich, CH-8093 Z\"urich, Switzerland} \affiliation{$^2$Walther-Meissner-Institut, Bayerische-Akademie der Wissenschaften,  D-85748 Garching, Germany} \affiliation{$^3$Geballe Laboratory for Advanced Materials
and Department of Applied Physics, Stanford University, Stanford, California
94305-4045, USA and Stanford Institute for Materials and Energy Sciences, SLAC National Accelerator Laboratory, 2575 Sand Hill Road, Menlo Park, California 94025, USA}

\date{\today}

\begin{abstract}
The two-dimensional rare-earth tri-tellurides undergo a unidirectional charge-density-wave (CDW) transition at high temperature and, for the heaviest members of the series, a bidirectional one at low temperature. Raman scattering experiments as a function of temperature on DyTe$_3$ and on LaTe$_3$ at 6 GPa provide a clear-cut evidence for the emergence of the respective collective CDW amplitude excitations. In the unidirectional CDW phase, we surprisingly discover that the amplitude mode develops as a succession of two mean-field, BCS-like transitions in different temperature ranges.  
\end{abstract}



\maketitle

Electronic instabilities are at the origin of phenomena as diverse as the formation of spin and charge density waves or superconductivity. Besides determining the ground state of the quantum system in which they occur, they also fundamentally affect its excitation spectrum. In particular, they induce new types of collective behavior, the investigation of which can be used to identify the kind of electronic order involved. In the charge density wave (CDW) state, on which we focus here, a gap opens up in the single particle spectrum, and two new collective modes, associated with the oscillations of the amplitude and of the phase of the CDW, respectively, appear \cite{grunerbook}. The oscillations of the phase involve a displacement of the electronic charge distribution with respect to the ionic positions, and consequently this mode is optically active. Such displacements do not occur for amplitude fluctuations and therefore the amplitude mode is expected to be Raman active \cite{grunerbook}. 

The paradigm of CDW forming materials are the quasi one-dimensional compounds  \cite{peierls}, the properties of which are nicely summarized in Ref. \onlinecite{grunerbook}. But electronically driven CDW states were also found and thoroughly investigated in novel two-dimensional (2D) layered compounds \cite{wilsonADVPHYS,rouxel,vuletic,kivelsonRMP,snow}, an effort motivated in part  by the fact that high temperature superconductivity in the copper-oxide systems may indeed emerge from a peculiar charge-ordering through the tuning of relevant parameters \cite{perali, kivelsonRMP}. A family of layered compounds which have attracted a lot of attention recently are the rare-earth ($R$) tri-tellurides $R$Te$_3$, first studied by DiMasi et al.  \cite{dimasi}. They host an $unidirectional$, incommensurate CDW already well above room temperature for all $R$ elements lighter than Dy \cite{ru1,ru2}, while in the heavy rare-earth tri-tellurides (i.e., $R$=Tm, Er, Ho, Dy) the corresponding transition temperature, $T_{CDW1}$, lies below $\sim$300 K and decreases with increasing $R$ mass. In the latter systems,  a further transition to a $bidirectional$ CDW state occurs at $T_{CDW2}$, ranging from 180 K for TmTe$_3$ to 50 K for DyTe$_3$ \cite{ru1,ru2}. The drastic change in transition temperatures with the size of the $R$ ion or externally applied pressure on a given material \cite{sacchettiESRF} is accompanied by a similarly large change in the properties of the CDW itself. In particular, the CDW gap of $R$Te$_3$ progressively collapses when the lattice constant is reduced, which, in turn, induces a transfer of spectral weight into the metallic component of the excitation spectrum  \cite{sacchetti1,sacchetti2,lavagnini}, the latter resulting from the fact that the Fermi surface in these materials is only partially gapped by the formation of the CDW. The response of this residual metallic component completely screens all optically active modes (including the collective CDW phase excitation) and makes their observation by infrared absorption methods impossible. This is why we turned to Raman scattering in our recent study of the influence of chemical and applied pressure on the lattice vibrational excitations in the lighter rare-earth tri-tellurides \cite{lavagniniRaman}. By combining experimental observations and numerical simulations, we were able to demonstrate the tight coupling between the CDW gap and the lattice degrees of freedom and to make a robust prediction for the Kohn anomaly inducing the CDW phase transition. An \emph{ab initio} calculation of the phonon dispersion relation for the orthorhombic pseudotetragonal ($a=c$) structure of LaTe$_3$ showed that two optical branches have an instability in the vicinity of the wave vector $q$=(2/7)$c^*$ with $c^*=\frac{2\pi}{c}$ \cite{lavagniniRaman}. The analysis of the associated atomic displacements shows that they are, in both cases, predominantly confined to the Te sheets and of transverse in-plane nature. The same picture is obtained for phonons
propagating in the $a$ direction.

While an unambiguous identification of the amplitude mode in $R$Te$_3$ by Raman scattering has been elusive so far, femtosecond pump-probe (FSPP) experiments, both in time- and angle-resolved photoemission on TbTe$_3$ \cite{shen} and in optical spectroscopy on $R$Te$_3$ ($R$= Tb, Dy and Ho)  \cite{yusupov} have provided rather convincing evidence for its existence in the unidirectional CDW phase. Although these latter experiments were performed down to temperatures well below  $T_{CDW2}$, the amplitude mode associated with the bidirectional CDW was not observed. 

Here, we present new Raman scattering investigations as a function of temperature on DyTe$_3$ at ambient pressure ($T_{CDW1}$=307 K, $T_{CDW2}$=49 K \cite{ru2})  and on LaTe$_3$ at 6 GPa, with a lattice constant between that of DyTe$_3$ and HoTe$_3$, and $T_{CDW1}\sim$ 260 K \cite{ru1,sacchettiESRF}. The covered spectral range extends from 5 to 200 cm$^{-1}$, i.e. well beyond the previous low frequency limit of about 65 cm$^{-1}$ \cite{lavagniniRaman}, thus giving access to the energy region in which the collective modes are expected to appear. In the temperature interval between $T_{CDW1}$ and $\sim$ 50 K, we observe that, in both systems, the amplitude mode in the unidirectional CDW phase exhibits a transition to a higher frequency around $T_{tr}\sim$ 170 K, which manifests itself in the form of a transfer of spectral weight between two peaks, whose positions as a function of temperature are well described by a BCS model \cite{grunerbook} for two condensate densities with different critical temperatures. Furthermore, a new feature below $\sim$ 50 K appears in DyTe$_3$. From its temperature dependence, it can be unambiguously identified with the amplitude mode associated with the bidirectional CDW phase.

The single crystalline samples of $R$Te$_3$ were grown by slow cooling of a binary melt \cite{ru1,ru2}. Raman spectra were collected on cleaved [010] surfaces in backscattering geometry. The Ar$^+$ and Kr$^+$ laser lines at 514 and 531 nm, respectively, were used for excitation. The heating in the 30 $\mu$m wide spot was kept below 5 K. Depending on the experimental requirements the scattered light was analyzed with a triple Jobin-Yvon T64000 and a double Jarrell-Ash 25-100 monochromator. We selected the polarization of the scattered light parallel to the incident one. High pressure was generated by means of a homemade diamond-anvil cell (DAC) with nitrogen as a pressure-transmitting medium. The pressure was determined by a standard ruby fluorescence technique \cite{mao}. 

\begin{figure}[!tb]
\center
\includegraphics[width=8.5cm]{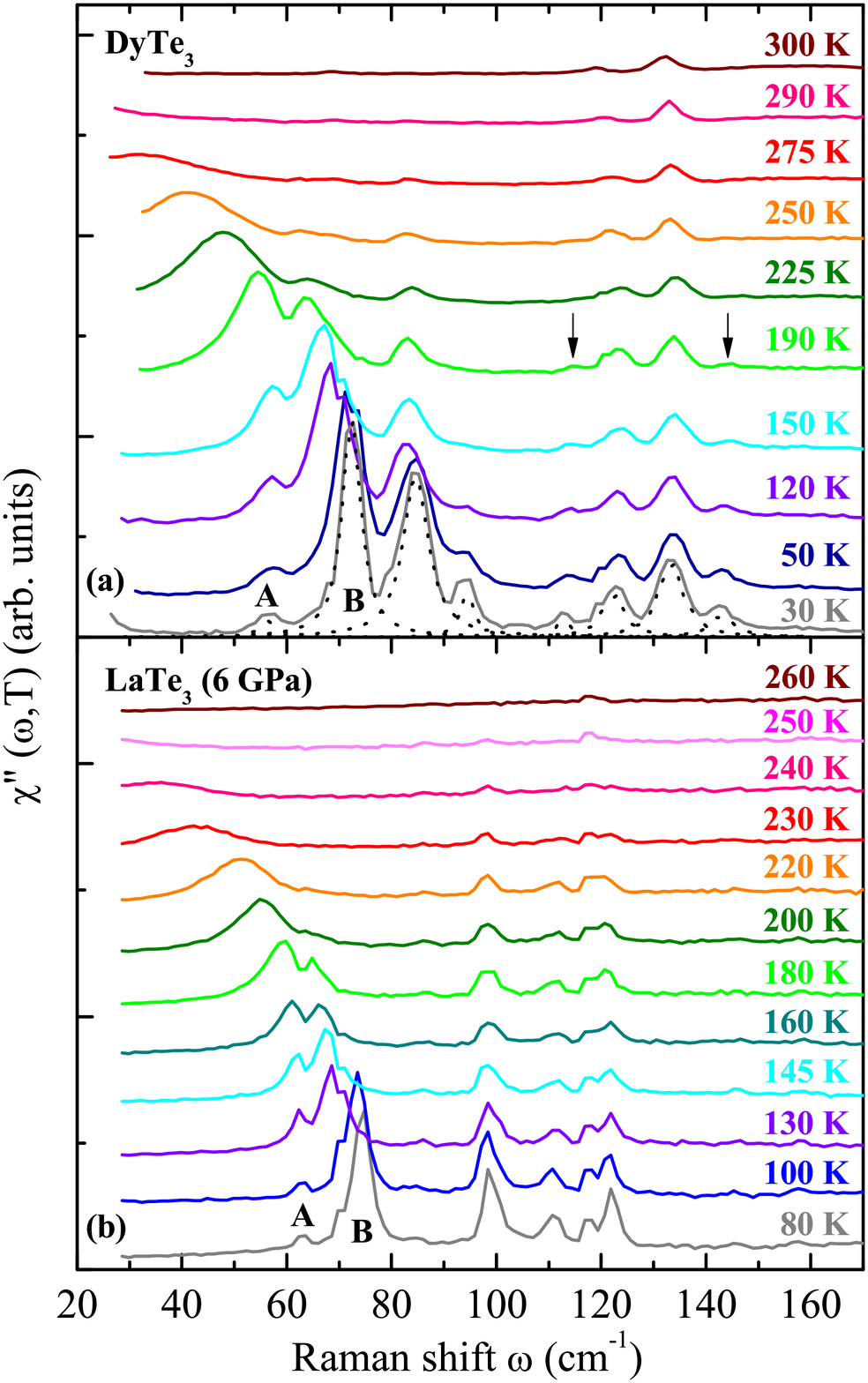}
\caption{(color online) Temperature dependence of the Raman scattering spectra of DyTe$_3$ at ambient pressure (a) and of LaTe$_3$ at 6 GPa (b). The spectra have been shifted for clarity. In panel (a) the oscillators employed for the data fits are shown for the measurement at 30 K and the arrows mark the weak features at 113 and 143 cm$^{-1}$. The labels A and B denote the weak mode at $\sim$ 60 cm$^{-1}$ and the sharp one at $\sim$ 70 cm$^{-1}$ at low temperatures, respectively.} \label{Experimental}
\end{figure}

Figure 1 displays the imaginary part $\chi''(\omega,T)$ of the Raman response obtained by dividing the measured spectra by the thermal Bose factor. The general trend in temperature for both sets of data is quite similar, suggesting that by hydrostatically compressing the lattice (as for LaTe$_3$ at 6 GPa) one can recover the properties and responses of chemically compressed $R$Te$_3$ (i.e., through substitution of the $R$ element). This confirms our earlier findings at 300 K as a function of pressure \cite{lavagniniRaman}, and is further supported by our data collected on HoTe$_3$ as a function of temperature, which will be presented elsewhere.

At low temperatures and deep in the CDW state, we observe well distinct and rather sharp modes. Above 80 cm$^{-1}$ the modes, previously ascribed to the Raman active phonons with $A_{1g}$, and $A_1$ and $B_1$ symmetry for the undistorted  (pseudotetragonal) and distorted phase respectively \cite{lavagniniRaman}, lose spectral weight with increasing temperature, while their width marginally changes as a function of temperature. 

For $\omega\le$ 80 cm$^{-1}$ and at low temperatures there is a weak peak close to 60 cm$^{-1}$ and a sharp one around 70 cm$^{-1}$ (labels A and B in Fig. 1), in agreement with the findings of Ref. \onlinecite{yusupov}. Their temperature dependence is specifically emphasized for DyTe$_3$ in Fig. 2a and 2b. Upon destroying the CDW state with increasing temperature the sharp B mode first softens, gets progressively broader and loses spectral intensity in favor of the A feature (Fig. 1a and 2b). At $T_{tr}$ of about 170 (160) K for DyTe$_3$ (LaTe$_3$ at 6 GPa), the two modes roughly share the same amount of spectral intensity  (Fig. 1 and 2(a-b)). Upon approaching $T_{CDW1}$ the energy of the B mode becomes constant, while its intensity drops above $T_{tr}$ and vanishes at $T_{CDW1}$, as does that of the A mode, whose resonance frequency saturates at 23 cm$^{-1}$ (Fig. 2a). In contrast mode C, which is only seen at high resolution (Fig. 2(a-b)), exists already above $T_{CDW1}$ and survives the transition, without displaying any temperature dependence. 

In the low temperature spectra of DyTe$_3$ an additional mode (D) in the range below 50 cm$^{-1}$ (Fig. 2c) is found, which disappears upon increasing the temperature above $T_{CDW2}$. Interestingly, its energy at $T_{CDW2}$ saturates at the same value (23 cm$^{-1}$) as the one of the A mode at $T_{CDW1}$, which we interpret as being due to an impurity scattering rate of this order of magnitude in our sample \cite{devereaux}. As a consequence, a quantitative analysis of the low energy spectral range in terms of Lorentz oscillators is not possible. Yet, from the temperature dependence observed in Fig. 2c, we can safely conclude that this mode is the collective amplitude mode of the bidirectional CDW state \cite{comment}.

\begin{figure}[!tb]
\center
\includegraphics[width=8.5cm]{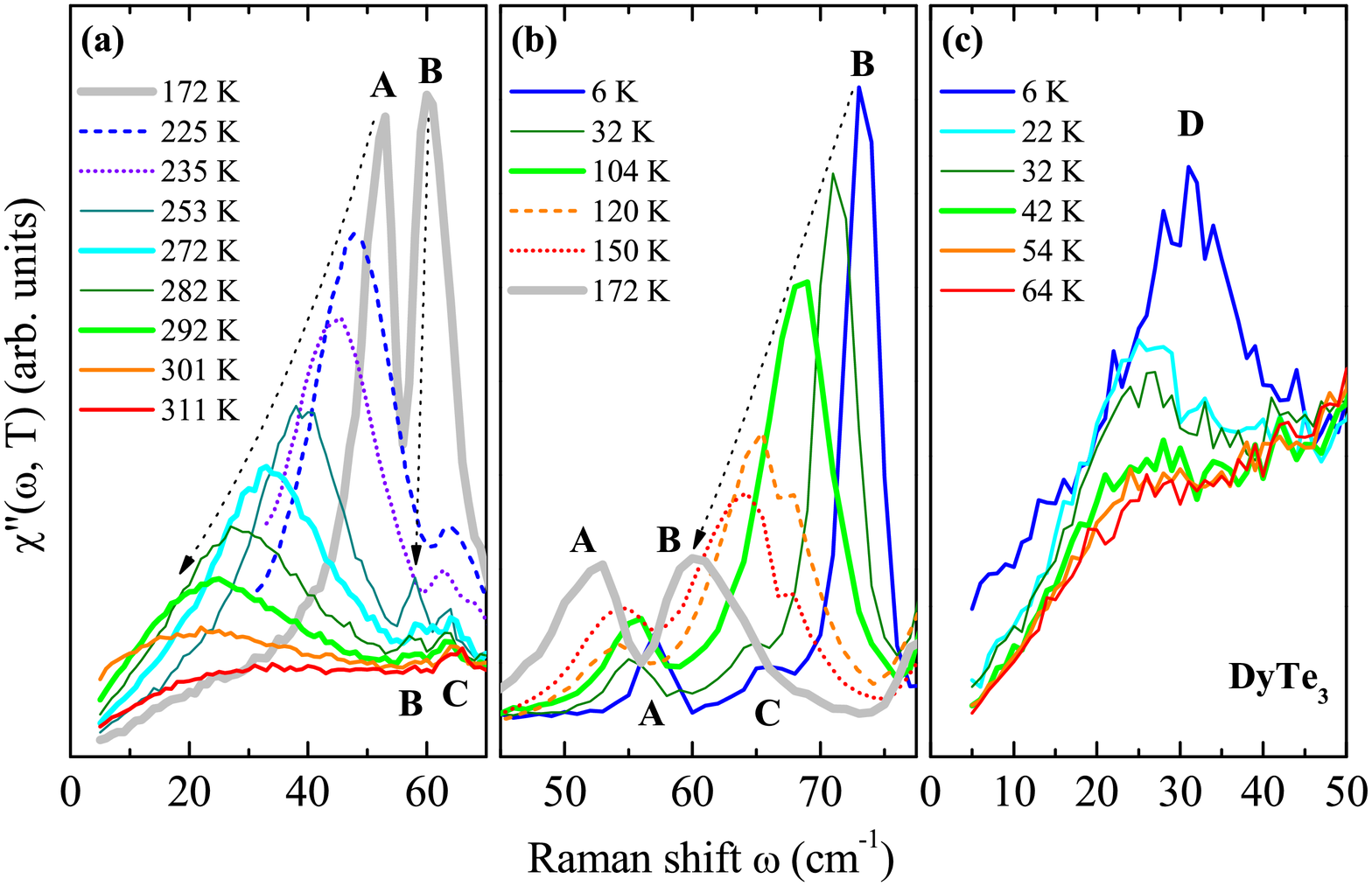}
\caption{(color online) Detailed view of the temperature dependence of the A, B (see Fig. 1a) and C modes of DyTe$_3$ above (a) and below (b) $T_{tr}$. The arrows highlight the trend of modes A and B with increasing temperature. (c) Low frequency interval characterized by mode D for temperatures close to and below $T_{CDW2}$. The relative intensity ratio among panels (a-c) is 2:8:1.} \label{Experimental}
\end{figure}

A quantitative analysis is possible for the modes evolving with the unidirectional CDW below $T_{CDW1}$. To this end, we perform a fit of the response function $\chi''$ with a series of damped harmonic oscillators \cite{Ramanfit}, after having subtracted a smooth background. A total of eight oscillators is employed for frequencies smaller than 170 cm$^{-1}$ and a sample of all fit components is shown in Fig. 1a. Figure 3 emphasizes the temperature dependence of the resonance frequencies and integrated intensities extracted from our fits for the two low-frequency modes A and B, for which we propose the following scenario.

\begin{figure}[!tb]
\center
\includegraphics[width=8.5cm]{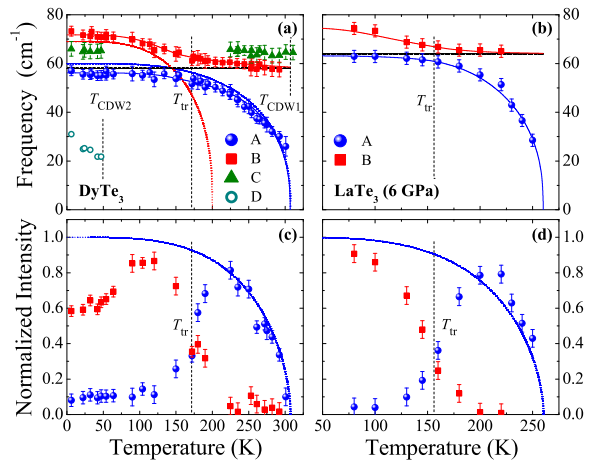}
\caption{(color online)  Temperature dependence of the resonance frequencies (a-b) and of the normalized integrated intensities (c-d) of the A (blue dot) and B (red square) modes of DyTe$_3$ at ambient pressure and of LaTe$_3$ at 6 GPa.  The dashed lines at about 59 (63) cm$^{-1}$ in panels (a-b) are the respective renormalized phonons which couple to the collective excitation. Panel (a) also reports the resonance frequencies of the C lattice mode (green triangle) and of the amplitude mode (D) for the bidirectional CDW of DyTe$_3$ (Fig. 2c). The blue and red lines in panels (a-b) are the fits of the resonance frequencies for modes A and B (see text). The thin dotted lines in panels (a) and (c-d) are the BCS predictions for the resonance frequencies ($\omega\sim n_c(T)$) and for the order parameter ($I_A(T)\sim\Delta(T)$) \cite{grunerbook,tinkham}, respectively. The normalization factor in panels (c-d) corresponds to the total intensity of both modes.} \label{fitfreq}
\end{figure} 

At $T_{CDW1}$ the system undergoes a transition into one of the two predicted unidirectional CDW states \cite{lavagniniRaman}. This transition results in a strong renormalization (from $\sim$ 120 cm$^{-1}$ to $\sim$ 60 cm$^{-1}$) of the frequency of the phonon at $q_{CDW1}$ of the second branch expected to soften according to the calculations of Ref. \onlinecite{lavagniniRaman}. As the temperature is lowered, the minimum on the free energy surface corresponding to the first unidirectional CDW moves towards smaller values of $q$, until it reaches a (saddle-) point at a temperature close to $T_{tr}$ where it becomes more favorable for the system to settle into the second calculated soft mode \cite{comment2}. The minute change in $q$-vector involved in this step does not alter the size of the gapped area on the Fermi surface (FS) and consequently it should not be seen in the electrical resistivity, especially in view of the relatively small effect induced by the drastic modification in FS topology occurring at $T_{CDW1}$ and $T_{CDW2}$ \cite{ru2}. What it does, however, is to interchange the original amplitude mode and renormalized phonon, so that feature A is an amplitude mode between $T_{CDW1}$ and $T_{tr}$ and a phonon below $T_{tr}$, while the opposite holds true for the B mode. The higher limiting frequency at zero temperature of the amplitude mode developing below $\sim T_{tr}$  can be explained as follows. On the one hand, a smaller wave-vector implies a higher frequency for the relevant phonon in the absence of CDW-transition according to Fig. 3b of Ref. \onlinecite{lavagniniRaman}; on the other hand it is plausible that the electron-phonon coupling constant, taken to be $q$-independent in conventional treatments \cite{grunerbook}, slightly increases with decreasing wave-vector when two close-by instabilities are involved. That the coupling increases below 200 K is also supported by the appearance of further two weak lines at 113 and 143 cm$^{-1}$ (arrows in Fig. 1a). The above situation can be quantitatively described by a model involving two temperature-dependent modes coupled to a phonon with a temperature-independent frequency. The results of Fig 3(a-b) were obtained by assuming that the two modes have resonance frequencies proportional to the respective condensate densities ($n_c(T)$, thin dotted lines in panel a), and that they interact with the  phonon at about 59 (63) cm$^{-1}$ for DyTe$_3$ (LaTe$_3$) (dashed line in Fig. 3(a-b)) with a coupling constant of $\sim$ 5 (4) cm$^{-1}$ \cite{note}. In standard CDW systems, where only one phonon branch is unstable, the frequency of the amplitude mode is proportional to $\sqrt{n_c}$ \cite{grunerbook}. The dependence found here confirms that the two unstable phonon branches are intimately coupled and cannot be considered separately as far as their frequencies are concerned. Our semi-quantitative model is similar to that previously used by Yusupov et al. \cite{yusupov}, based however on the Ginzburg-Landau temperature dependence of the amplitude mode. This latter approach suffers from a few weakness: namely, the resulting temperature-dependence of the amplitude mode frequency is not consistent with the mean-field prediction and the model itself would be only valid at $T \sim T_{CDW1}$ and clearly fails to reproduce our high resolution data for $T\ll T_{CDW1}$.
 
Our model implies a transfer of spectral intensity among A and B: the intensity of mode A first rises below $T_{CDW1}$ with decreasing temperature, around $T_{tr}$ domains characteristic of the second calculated distortion start to form, leading to a transfer of spectral intensity to feature B at the cost of the high temperature (A) one (Fig. 3(c-d)). Close to $T_{CDW1}$ there is a good agreement between the integrated intensity of feature A and the prediction for the BCS order parameter (i.e., $I_A(T)\sim\Delta(T)$) \cite{tinkham,comment3}. Another peculiarity is the loss of spectral intensity of the B mode (Fig. 3c) at low temperatures ($T\lesssim$ 50 K), which agrees with the increase of intensity observed below $T_{CDW2}$ for the collective mode associated with the bidirectional CDW (Fig. 2c). 

In summary, we have clearly identified the collective amplitude modes of both uni- and bidirectional CDW ground states in $R$Te$_3$. Upon entering the unidirectional CDW state with decreasing temperature, we have observed an astonishing transfer of spectral intensity between two excitations, giving clear evidence for a two-step transition at $T_{CDW1}$ and close to $T_{tr}$. This cascade of transitions is rather unconventional but yields a consistent picture of the evolution of the collective excitations in $R$Te$_3$ over the whole range of existence of the CDW.\\

The authors wish to thank A. Sacchetti and D. Einzel for fruitful discussions.
This work has been supported by the Swiss National Foundation for the Scientific Research and by the NCCR MaNEP pool, as well as by
the Department of Energy, Office of Basic Energy Sciences under contract
DE-AC02-76SF00515. The work at Garching was partially supported by the Deutsche Forschungsgemeinschaft via the projects Ha2071/3-4 and Ha2071/5-1. Grant Ha2071/3-4 is part of Research Unit FOR538.


\begin{thebibliography}{99}
\bibitem{comment} An equivalent experiment at low frequencies and temperatures was not performed on LaTe$_3$ at 6 GPa.
\bibitem{note} The frequency of the renormalized phonon and the latter's coupling constant
to the amplitude mode(s) were taken equal above and below $T_{tr}$, as they cannot be
disentangled within this simple model.
\bibitem{comment3} The intensities of the modes above 80 cm$^{-1}$ do not scale linearly with the BCS order parameter, in accord with our pressure dependent experiment \cite{lavagniniRaman}.
\end{thebibliography}
\end{document}